\documentclass[prl,twocolumn,showpacs,showkeys,amssymb,floatfix]{revtex4}
\usepackage{graphicx}
\usepackage{dcolumn}
\usepackage{bm}
\begin{document}
\title{Nondispersive and dispersive collective electronic modes in 
carbon nanotubes}
\author{Ricardo Perez}
\email{rtperez@scs.ryerson.ca}
\author{William Que}
\email{wque@ryerson.ca}
\affiliation{Department of Physics, Ryerson University, 350 Victoria Street,
Toronto, Ontario, Canada M5B 2K3}
\begin{abstract}
We propose a new theoretical interpretation of the electron energy-loss spectroscopy 
results of Pichler {\it et al.} on bulk carbon nanotube samples. The 
experimentally found nondispersive modes have been attributed by Pichler {\it et al.} to interband 
excitations between localized states polarized perpendicular to the nanotube axis.  This interpretation has been challenged by a theorist who attributed the modes to optical plasmons carrying nonzero angular 
momenta. We point out that both interpretations suffer from difficulties. From our theoretical results of the loss functions for individual carbon 
nanotubes based on a tight-binding model, we find that the nondispersive modes could be due to collective electronic modes in 
chiral carbon nanotubes, while the observed dispersive mode should be due to 
collective electronic modes in armchair and zigzag carbon nanotubes. 
Momentum-dependent electron energy-loss experiments on individual carbon 
nanotubes should be able to confirm or disprove this interpretation decisively.
\end{abstract}

\pacs{71.20.Tx, 78.67.Ch, 73.20.Mf, 71.10.Pm}

\keywords {Carbon nanotubes, plasmons, EELS, collective modes}

\maketitle


Carbon nanotubes have been on the centre stage of physics research for over a 
decade, for good reasons. Other than the long list of practical applications 
possible, the fundamental physical properties of carbon nanotubes have been 
extremely interesting and challenging. Notably, metallic carbon nanotubes have 
been found to exhibit Luttinger liquid behavior \cite{Bockrath}, and whether a 
carbon nanotube is metallic or semiconducting is dependent on the chirality of 
the tubes. In a Luttinger liquid, it is well known that single-particle 
excitations are suppressed, thus the collective electronic modes or plasmons 
play an extremely important role in carbon nanotubes.

Momentum-dependent electron energy loss spectroscopy (EELS) as carried out by 
Pichler {\it et al.} \cite{Pichler,Knupfer,Liu} offers an excellent tool for 
studying plasmons in carbon nanotubes. Their experiment was performed first 
on bulk samples of single wall carbon nanotubes \cite{Pichler,Knupfer} and 
later on magnetically aligned bundles of single wall carbon nanotubes 
\cite{Liu}. In the low energy range of the spectrum, the experimental findings 
are: (1) a dispersive mode as function of momentum transfer in 
the $5-8$ eV range;  (2) several nondispersive modes at lower energies. The 
dispersive mode was attributed to the $\pi$-plasmon without controversy. As 
for the nondispersive modes, no theory predicted their existence, and 
Pichler {\it et al.} interpreted them  in terms of interband excitations between 
localized states polarized perpendicular to the nanotube axis.

Three years later, theorist Bose \cite{Bose} challenged this interpretation, 
noting that according to EELS theory \cite{Raether}, the experiment should 
measure the collective electronic modes. Based on a plasmon calculation 
using a model of free electron gas confined to a cylindrical surface, he  
suggested an alternative interpretation of the nondispersive modes in terms 
of optical plasmons carrying nonzero angular momenta.  However, a close 
inspection of the calculated plasmon dispersion curves presented in an earlier 
paper by Longe and Bose \cite{Longe} reveals difficulties with this 
interpretation. In Fig. 1 of that paper, one can see that the acoustic plasmon 
which carries zero angular momentum is the lowest in energy and most 
dispersive. Plasmons with nonzero angular momenta are all optical, and as 
the angular momentum increases, the energy increases and the amount of 
dispersion decreases. While Bose did not clarify if the dispersive mode 
corresponds to a zero angular momentum mode or not, difficulties arise 
regardless of how the dispersive mode is assigned: if it is assigned as a 
zero angular momentum mode, the optical plasmons should have {\it higher} 
energies than the dispersive mode, not at lower energies as experimentally 
observed; if it is assigned as a nonzero angular momentum mode, for the 
energies to be in correct order, it must have larger angular momentum than the 
nondispersive modes, but larger angular momentum should correspond to less 
dispersion!

The bulk sample used in the experiment of Pichler {\it et al.} had a mean 
diameter of $1.4$ nm, and nondispersive modes were observed at $0.85$, $1.45$, 
$2.0$, and $2.55$ eV. Optical absorption measurements by Jost {\it et al.} 
\cite{Jost} on carbon nanotube containing-soot revealed excitations at $0.72$, 
$1.3$, and $1.9$ eV for the mean diameter of $1.29$ nm. Since the gaps between 
van Hove singularities in the electronic density of states is known to be 
inversely proportional to the diameter, single-particle excitation energies 
should be larger in smaller diameter carbon nanotubes. However, the observed 
excitations in the experiment of Jost {\it et al.} appear to be at smaller 
energies compared to those observed in the experiment of Pichler {\it et al.} 
To reconcile the two experiments, one has to assume that the nondispersive 
modes observed by Pichler {\it et al.} are collective rather than 
single-particle modes. A more recent paper by Liu {\it et al.} \cite{Liu2} 
comparing optical absorption with EELS suggests that the nondispersive modes 
in the EELS are collective excitations caused by the optically allowed 
transitions. This could be a viable interpretation (barring the perpendicular polarization), making these modes analogous to 
the intersubband plasmons in quantum wires \cite{Quew1,Quew2,Sarma}. However, recent experimental and theoretical results \cite{Islam} on polarized
 optical absorption of aligned single wall carbon nanotubes of 1.35 nm in average diameter show that when the light is polarized parallel to the tube axis, the absorption spectra have several peaks below 3 eV, but when the light is polarized perpendicular to the tube axis, the absorption spectra become essentially featureless. Similar results were obtained earlier for tubes of much smaller diameter (0.4 nm) \cite{ZMLi,WLiang}. These results cast doubt on the interpretation of the nondispersive modes in terms of excitations polarized perpendicular to the tube axis. 
Clearly,
over six years after the initial discovery of the nondispersive modes in EELS, the origin of the modes remains a puzzle.

In this paper, we present our theoretical results on the loss functions of 
individual carbon nanotubes, and shed some light on the origin of the 
nondispersive modes. In particular, we propose that the nondispersive modes 
are inter(sub)band plasmons from chiral carbon nanotubes which have small 
Brillouin zones. These collective modes generally are not polarized perpendicularly to the tube axis. Further experiments are suggested to decisively determine 
the validity of this interpretation.

\begin{figure}[t]

\begin{center}

\includegraphics[width=.75\linewidth,angle=-90]{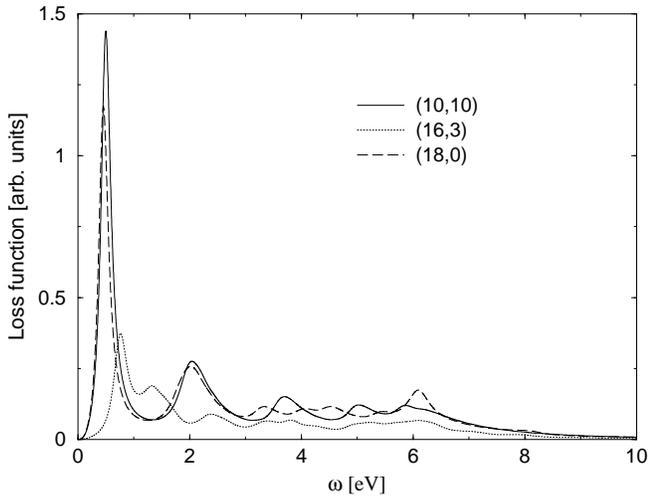}

\caption{\label{fig1} Loss function {\it Im}($-1/\epsilon_M(q_0,L,\omega)$)
computed for $q_0=0.04$ $\AA^{-1}$ and zero angular momentum $(L=0)$ for tubes 
with different chirality but similar radius about $7$ $\AA$. The $(10,10)$ 
armchair tube and the $(18,0)$ zigzag tube are both metallic. The $(16,3)$ 
chiral tube is semiconducting.}
\end{center}

\end{figure}

It is well known that the electronic properties of carbon nanotubes are 
dependent on the chirality. Whether a $(n,m)$ carbon nanotube is metallic or 
not depends on if the difference $n-m$ is divisible by $3$. Such important 
details are not captured by a free electron gas type model. On the other hand, 
a tight-binding model \cite{Saito} is known to produce the electronic band 
structures of carbon nanotubes very well as long as the radius is not too 
small. We use such a tight-binding model for $\pi$ band electrons to study the 
collective electronic excitations of individual carbon nanotubes. The 
theoretical framework is the well-used random phase approximation (RPA) 
theory, which has been applied successfully to many systems including 
quantum wires \cite{Quew1,Quew2,Sarma}. While this theory is usually used 
for Fermi liquids, Li, Das Sarma, and Joynt \cite{Li} have shown that for 
a quantum wire with only one occupied subband, this theory gives the 
correct result for a Luttinger liquid. More recently, Que \cite{Que} has 
applied this theory to metallic carbon nanotubes, and obtained the same 
results as other established methods for studying Luttinger liquids. Based 
on these findings, it was concluded that the RPA theory is suitable for 
studying plasmons in both Fermi liquids and Luttinger liquids.

\begin{figure}[t]

\begin{center}

\includegraphics[width=.75\linewidth,angle=-90]{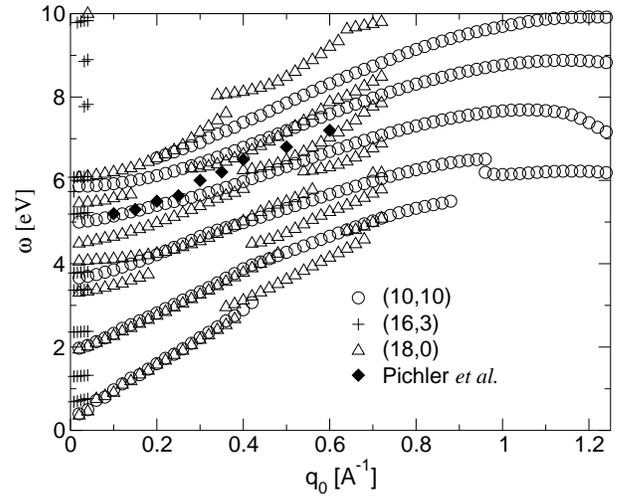}

\caption{\label{fig2} Dispersion curves for the collective electronic modes 
with angular momentum index $L=0$, for the same three carbon nanotubes as in
Fig. 1. The solid diamonds are experimental results on the dispersive mode 
from Pichler {\it et al.} The Brillouin zone edges for the $(10,10)$, 
$(18,0)$, and $(16,3)$ carbon nanotubes are at $q_0=1.26$, $0.73$, and 
$0.041$ $\AA^{-1}$, respectively.} 
\end{center}

\end{figure}

\begin{figure}[!]

\begin{center}

\includegraphics[width=.75\linewidth,angle=-90]{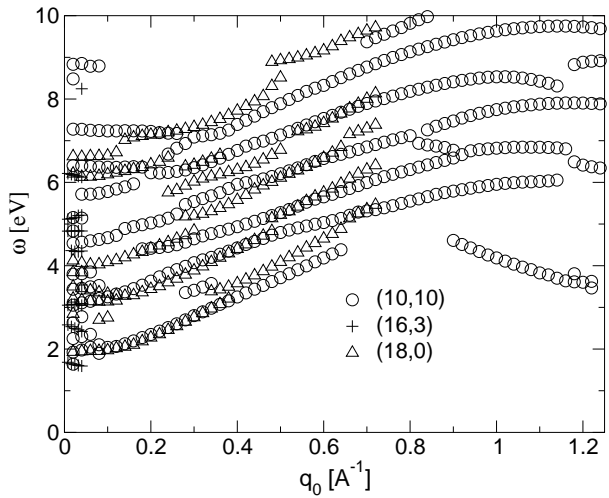}

\caption{\label{fig3} Dispersion curves for the collective electronic modes 
with angular momentum index $L=1$, for the same three carbon nanotubes as in 
Fig. 1.}
\end{center}
\end{figure}

Fig. \ref{fig1} shows the loss functions of the $(10,10)$ armchair carbon 
nanotube (radius $R=6.88 \AA$), the $(18,0)$ zigzag carbon nanotube 
 ($R=7.15 \AA$), and the $(16,3)$ chiral carbon nanotube ($R=7.02 \AA$). The 
angular momentum is a good quantum number, and only the zero angular momentum 
modes are shown. Each loss function has several peaks but becomes featureless 
beyond $12$ eV ($\sigma$ band electrons are not included in the model).

By scanning the loss functions to find peak positions at different 
wavevectors, we produce the dispersion curves of the collective 
electronic modes in Figs. \ref{fig2} and \ref{fig3}, for wavevectors along 
the tube axis up to the Brillouin zone edge of the corresponding carbon 
nanotube. Assuming a carbon-carbon bond length of $a_{C-C}=1.44$ 
$\AA$, it can be shown that all armchair carbon nanotubes have the same 
Brillouin zone edge of $\pi/T=1.26$ $\AA^{-1}$ ($T$ is the length of the 
translational vector \cite{Saito}), and all zigzag carbon nanotubes have the 
same Brillouin zone edge of $\pi/T=0.73$ $\AA^{-1}$, but different chiral 
carbon nanotubes have different Brillouin zone sizes. Those $(n,m)$ chiral 
nanotubes for which the greatest common divisor among $2n+m$ and $2m+n$ is $1$ 
have the smallest Brillouin zones, with $\pi/T=\pi/(3a_{C-C}\sqrt{n^2+m^2+nm})$. For 
the $(16,3)$ chiral carbon nanotube, its Brillouin zone edge is at 
$\pi/T=0.041$ $\AA^{-1}$. 
Some of the curves for the $(10,10)$ and $(18,0)$ carbon nanotubes terminate 
before reaching the Brillouin zone edge due to vanishingly small peak 
amplitudes. Clearly, the $(10,10)$ armchair tube and the $(18,0)$ 
zigzag tube both have dispersive modes for all the computed $L$, and we find 
this to be generally true for armchair and zigzag tubes. On the other hand, 
the collective electronic modes of the $(16,3)$ chiral tube have little 
dispersion, and so do many other chiral tubes. The reason for the lack of dispersion is the much smaller 
Brillouin zone.

If we compare the results in Figs. \ref{fig2} and \ref{fig3} with the results 
of Longe and Bose \cite{Longe}, a major difference is that in the latter, 
there is only one branch of collective mode for each angular momentum index 
$L$, while in our results, we find many branches for each angular momentum 
index. This is due to the band structures of carbon nanotubes with many 
occupied and many empty (sub)bands. Generally speaking, when $L$ is increased, 
excitation energies increase, and dispersion is reduced. These qualitative 
features are already present in the free electron gas type model. Unlike Bose, 
we find there is no need for the nonzero angular momentum modes in order to 
explain the nondispersive modes. Since the experiment of Pichler {\it et al.} 
was performed on bulk samples ($7$ $\AA$ mean radius), the measured 
spectra contain contributions from many carbon nanotubes of different 
chirality. The nondispersive modes could be due to chiral carbon nanotubes, 
and the dispersive mode should be due to armchair and zigzag carbon nanotubes. Experimentally, only one dispersive mode was found, but since the peak 
of the dispersive mode was a couple eV broad, it is possible that several 
modes of large amplitude contributed to the broad peak.

Since intertube coupling shifts the energies of the collective electronic 
modes higher \cite{Que2}, it is not possible to match the calculated energies 
in this work for individual carbon nanotubes to experimental results on bulk 
samples where intertube coupling is present. To allow an exact comparison 
between theory and experiment, it is desirable to obtain momentum-dependent 
EELS for individual carbon nanotubes, and such experiments should determine 
decisively the validity of the interpretation offered in this paper. We note 
that Reed and Sarikaya \cite{Reed} have already done EELS work on individual 
carbon nanotubes (but not momentum-dependent measurements) and noticed 
variations in results from different tubes. Therefore momentum-dependent 
EELS for individual carbon nanotubes are not only desirable, but also 
achievable. If our prediction is confirmed experimentally, eventually EELS 
could become a potential tool for identifying the chirality of individual 
carbon nanotubes.


\begin{acknowledgments}
This work was supported by the Natural Sciences and Engineering Research 
Council of Canada. 
\end{acknowledgments}


\end{document}